\PassOptionsToPackage{colorlinks=true,citecolor=darkred,linkcolor=darkred,urlcolor=darkred,pdfborder={0 0 0}}{hyperref}

\newif\ifrevtexavailable
\IfFileExists{revtex4-2.cls}{%
  \revtexavailabletrue
  \documentclass[aps,prd,notitlepage,nofootinbib,superscriptaddress,floatfix]{revtex4-2}%
}{%
  \revtexavailablefalse
  \documentclass[11pt,a4paper]{article}%
}
\usepackage[normalem]{ulem}
\pdfoutput=1

\usepackage[utf8]{inputenc}
\usepackage[T1]{fontenc}
\usepackage[dvipsnames]{xcolor}
\usepackage{amsmath,amssymb,mathtools,bm}
\usepackage{booktabs,array}
\usepackage{graphicx}
\usepackage{hyperref}

\ifrevtexavailable\else
  \usepackage[margin=2.3cm]{geometry}
\fi

\definecolor{darkred}{rgb}{0.6,0,0}

\allowdisplaybreaks[2]
\newcommand{\ignore}[1]{}

\newcommand{\AddrIFIC}{Instituto de F\'isica Corpuscular (IFIC), CSIC, Universitat de Val\`encia, 46980 Paterna, Valencia, Spain}
\newcommand{\AddrIISER}{Department of Physics, Indian Institute of Science Education and Research Bhopal, Bhopal Bypass Road, Bhauri, Bhopal 462066, India}

\begin{document}

\title{\boldmath Escaping Solar Capture: Majoron mediated Elastic Dark Matter \\for the LZ 248~keV Event}

\ifrevtexavailable
\author{Salvador Centelles Chuli\'a}
\email{salcen@ific.uv.es}
\affiliation{\AddrIFIC}

\author{Ayan Chattaraj}
\email{ayan23@iiserb.ac.in}
\affiliation{\AddrIISER}

\author{Rahul Srivastava}
\email{rahul@iiserb.ac.in}
\affiliation{\AddrIISER}

\author{Surabhi Srivastava}
\email{surabhi24@iiserb.ac.in}
\affiliation{\AddrIISER}

\begin{abstract}
\vspace{0.2cm}
We interpret the $248~\mathrm{keV_{nr}}$ nuclear-recoil candidate reported by LZ through elastic dark matter scattering mediated by a light Majoron. The Majoron provides a natural origin for the required pseudoscalar interaction: it is the pseudo Nambu-Goldstone boson associated with spontaneous lepton number breaking, directly connecting the recoil signal to neutrino mass generation. After a model independent analysis of this idea,  we present a concrete realization in a scotogenic ultraviolet completion in which the lightest singlet Majorana fermion is dark matter and acquires both its mass and Majoron coupling from the symmetry breaking dynamics. A vector like up type quark and a type-I Dirac seesaw in quark sector communicate the Majoron interaction to the up quark. At low energies, the resulting portal generates the nonrelativistic operator $\mathcal O_6$, whose momentum dependence reproduces the high-recoil event without overpopulating the lower energy bins. Its spin structure removes the dominant heavy element solar capture channels, while capture on hydrogen is suppressed by poor kinematic matching and small momentum transfer. This construction simultaneously reproduces the LZ recoil spectrum and the observed thermal relic abundance while strongly suppressing solar capture.
\end{abstract}

\maketitle
\else
\author{Salvador Centelles Chuli\'a$^{1}$, Ayan Chattaraj$^{2}$, Rahul Srivastava$^{2}$, Surabhi Srivastava$^{2}$\\[2mm]
\small $^{1}$\AddrIFIC\\
$^{2}$\AddrIISER}
\date{}
\maketitle
\fi

\section{Introduction}

The LUX-ZEPLIN (LZ) Collaboration recently extended its search for dark matter interactions to nuclear recoil energies up to approximately $270~\mathrm{keV_{nr}}$ and reported one candidate event at $E_R=248\pm23_{\rm stat}\pm23_{\rm syst}~\mathrm{keV_{nr}}$, with a global background only significance of $2.6 \sigma$~\cite{LZ:2026highrecoil}. Although a single event cannot establish a dark matter (DM) signal, its position near the upper end of the analysis window is difficult to reconcile with conventional elastic DM scattering, which generically predicts a much larger event rate at lower recoil energies. An interpretation in terms of coherent elastic neutrino--nucleus scattering (CE$\nu$NS) faces the same difficulty: standard and non-standard astrophysical neutrino fluxes, neutrinos from dark matter annihilation or decay, and neutrinos from primordial black hole evaporation cannot reproduce the candidate without overpopulating the lower-energy bins or violating existing constraints~\cite{Chattaraj:2026fxn}. However, inelastic neutrino up-scattering can potentially explain the excess~\cite{Jeesun:2026vzo,Brdar:2026ukx}. These results disfavour  any potential within SM origin of the LZ event  and strengthen the motivation for a dark matter interaction with a non-standard recoil dependence.

Inelastic DM-nucleus scattering provides the most immediate kinematic mechanism for shifting the recoil spectrum towards high energies~\cite{Tucker-Smith:2001myb}. In endothermic scattering, the incident particle transitions into a heavier state, suppressing low-energy recoils and selecting the high-velocity tail of the Galactic distribution. Reproducing the LZ event typically requires a mass splitting of order $\delta\sim300~\mathrm{keV}$~\cite{Su:2026inelastic,DiMauro:2026kinematic,McCabe:2026seasonal,Dent:2026inelastic,Fan:2026inelastic,Ahmed:2026isotope,Palmisano:2026inference,He:2026idw}. This possibility has been explored in realizations based on Higgsino and electroweak dark matter, inert scalars, gauge portals and neutrino mass models~\cite{FanReece:2026higgsino,Freese:2026higgsino,Wu:2026higgsino,Visinelli:2026PQ,Smirnov:2026EWDM,WangXiao:2026IDM,Borah:2026singletDoublet,Kumar:2026lgi,Borah:2026ris}. Exothermic scattering and other proposed explanations based on boosted dark matter, transition moments and unconventional nuclear processes have also been explored~\cite{Dent:2026inelastic,deLima:2026exothermic,Baer:2026exothermic,Fan:2026inelastic,Palmisano:2026inference,Alhazmi:2026boosted,Kannike:2026boosted,Liang:2026boosted,Heikinheimo:2026boosted,He:2026dipole,Asadi:2026magnetic,AghaieStrumia:2026neutron,Uttayarat:2026neutron,LeeTakahashiTsai:2026neutronpair}.

Solar capture of DM places strong constraints on the inelastic interpretation of the event. DM particles falling into the Sun reach velocities well above those available in terrestrial detectors, opening endothermic transitions on heavy solar elements and producing a captured population whose annihilation products are constrained by neutrino telescopes~\cite{Gould:1987ir,Gould:1987ju,Catena:2015uha}. This mechanism was established for inelastic dark matter well before the LZ result~\cite{Nussinov:2009ft}. In the scotogenic model, de Boer et al.\ found in 2021 that scalar dark matter with mass splittings below approximately $500~\mathrm{keV}$ can generate large IceCube event rates even when terrestrial scattering is kinematically suppressed~\cite{deBoer:2021pon}. The splitting required by the LZ event therefore lies directly within the regime in which solar capture remains efficient. Following the LZ result, solar capture has also been shown to exclude the canonical thermal Higgsino interpretation~\cite{Pospelov:2026solar}, while more general analyses have emphasized the dependence on the nuclear interaction, DM mass, annihilation channels and evolution of the captured population~\cite{DiMauroShaikh:2026solar,Nguyen:2026solar,Ghosh:2026higgsino, Murayama:2026apt}.

A recent scotogenic interpretation proposed suppressing solar capture through an isospin-violating cancellation of the effective proton coupling, motivated by the large hydrogen abundance in the Sun~\cite{Okada:2026scotogenic}. For the mass splittings of a few hundred keV relevant to the LZ event, however, scattering on hydrogen is kinematically closed while heavier solar elements remain accessible. Cancelling the proton coupling therefore does not suppress the heavy element capture channels that control the rate in this regime~\cite{Nussinov:2009ft,deBoer:2021pon,Catena:2015uha,Pospelov:2026solar,DiMauroShaikh:2026solar}.

Elastic scattering offers a different route to the observed spectrum. Pseudoscalar couplings to both fermionic dark matter and nucleons generate the non-relativistic operator $\mathcal O_6$~\cite{Fitzpatrick:2012ix,Anand:2013yka}, whose spin and momentum dependence suppresses low-energy recoils and enhances their relative weight at high energies. Such interactions were studied in direct detection contexts well before the LZ result~\cite{Freytsis:2010ne,Dienes:2013xya}. Explicit realizations proposed for the LZ event include an axion portal~\cite{Unwin:2026axion} and a renormalizable $2{\rm HD}+a$ model~\cite{Arcadi:2026pseudoscalar}.

The same spin and momentum dependence also suppresses solar capture: the pseudoscalar-pseudoscalar interaction has no coherent response on spin-zero nuclei, while scattering on hydrogen is reduced by the small momentum transfer. A recent analysis found that the region favoured by LZ remains allowed by IceCube and Super-Kamiokande even for annihilation directly into neutrinos~\cite{Bose:2026solar}. Part of this region lies beyond the validity of the contact effective field theory description, motivating an explicit light mediator realization.

Light pseudoscalars arise naturally from spontaneous lepton number breaking~\cite{Chikashige:1980qk,Chikashige:1980ui,Gelmini:1980re}. The associated Nambu-Goldstone boson, the Majoron, acquires a mass in the presence of a small explicit breaking while retaining pseudoscalar couplings to the fermions whose masses originate from the symmetry breaking vacuum expectation value. This connects the interaction suggested by the LZ spectrum with models of neutrino mass. The link between radiative neutrino masses and dark matter in the field content now associated with the scotogenic model was first proposed by Tao in 1996 and independently developed by Ma in 2006~\cite{Tao:1996vb,Ma:2006km}.

In this work, we show how the Majoron can act as the mediator required to explain the LZ event. We first discuss this at the model independent effective operator level, showing the general construction and the allowed parameter space needed to explain the LZ event. We then construct a scotogenic Majoron realization of the pseudoscalar interaction favoured by the LZ high-recoil event. The dark matter candidate is the lightest singlet Majorana fermion $N_1$, while a single up-type vector-like quark communicates the Majoron interaction to the visible sector. A quark type-I Dirac seesaw generates the physical up quark mass and decouples the visible sector portal coupling from the Majorana dark matter coupling. We determine the parameter space in which the resulting $\mathcal O_6$ interaction reproduces the LZ event without overproducing lower-energy recoils, yields the observed thermal relic abundance and suppresses solar capture.

\section{Majoron mediated elastic scattering at high recoil and solar capture}

This section introduces the three ingredients underlying our construction. We first review how spontaneous lepton number breaking connects neutrino mass generation to the emergence of the Majoron. We then analyze, independently of any specific ultraviolet completion, elastic scattering of fermionic dark matter through a pseudoscalar mediator and its implications for the LZ high-recoil event. Finally, we discuss how the same spin and momentum dependence suppresses dark matter capture in the Sun. A concrete scotogenic realization in which the pseudoscalar mediator is identified with the Majoron is presented in the following section.

\subsection{Neutrino mass generation and the emergence of the Majoron}

We begin with a brief model independent review of neutrino mass generation and the natural emergence of a pseudoscalar, commonly known as the Majoron, in ultraviolet completions with spontaneous lepton number breaking~\cite{Chikashige:1980qk,Chikashige:1980ui,Gelmini:1980re,Schechter:1981cv,Schechter:1980gr}. Modern classifications of Majoron realizations in seesaw models can be found, for example, in Refs.~\cite{CentellesChulia:2024TypeI,CentellesChulia:2025eck}. Within the $SU(2)_L\otimes U(1)_Y$ electroweak theory, neutrino masses can be accommodated by introducing the dimension-five Weinberg operator~\cite{Weinberg:1979sa}, $\overline{L_i^c}L_jHH$, with $i,j=1,2,3$, where $L_i$ and $H$ denote the lepton and Higgs doublets, respectively. The Weinberg operator explicitly breaks the accidental global lepton number symmetry $U(1)_L$ to its $\mathbb Z_2$ subgroup.

Alternatively, neutrino masses can arise from the spontaneous breaking of lepton number. Introducing a scalar $\Phi$ that is a singlet under the Standard Model gauge group and carries two units of lepton number, a simple effective realization of spontaneous lepton number breaking is provided by the operator
\begin{equation}
\frac{\mathcal C_{ij}}{\Lambda^2}\,\overline{L_i^c}L_jHH\Phi .
\label{eq:weinberg}
\end{equation}
The $U(1)_L$ symmetry is spontaneously broken by the vacuum expectation value of $\Phi$, $\langle\Phi\rangle=v_\phi/\sqrt{2}$, while $\langle H\rangle=v/\sqrt{2}$ breaks the electroweak symmetry as in the Standard Model. After symmetry breaking, neutrinos acquire masses, with the light neutrino mass matrix given by
\begin{equation}
m_{ij}=\frac{\mathcal C_{ij}v^2v_\phi}{\Lambda^2}.
\end{equation}

The spontaneous breaking of the global lepton number symmetry also produces a Nambu-Goldstone boson, the Majoron $J$, as a physical degree of freedom. More generally, the spontaneous breaking of a global lepton number symmetry implies the presence of a Majoron, independently of the dependence of the neutrino mass scale on \(v_\phi\). This makes Majoron mediated interactions a natural possibility in neutrino mass models with spontaneous lepton number breaking. We investigate whether the Majoron can mediate interactions between dark matter and nucleons, thereby connecting neutrino mass generation, dark matter and direct detection through a common $U(1)_L$ symmetry. We first analyze the LZ $248~\mathrm{keV_{nr}}$ event without specifying the ultraviolet completion and subsequently present a concrete scotogenic realization.

\subsection{Elastic scattering at high recoil}
\label{subsec:elastic-high-recoil}
Before specifying the ultraviolet completion, it is useful to isolate the interaction responsible for the recoil spectrum and analyze it in a model independent way. Consider a fermionic dark matter particle $F$ and a pseudoscalar mediator $J$, with interactions
\begin{equation}
\mathcal L_{\rm eff}\supset-i g_FJ\overline F^c\gamma_5F-iJ\left(g_u\overline u\gamma_5u+g_d\overline d\gamma_5d\right).
\label{eq:general-portal}
\end{equation}
For the model developed in Sec.~\ref{sec:scoto}, $F$ will be identified with the lightest singlet Majorana fermion $N_1$. At the nucleon level we write
\begin{equation}
\mathcal L_{\rm eff}\supset -iJ\sum_{\mathcal N=p,n}g_{J\mathcal N}(q^2)\overline{\mathcal N}\gamma_5\mathcal N ,
\label{eq:nucleon-matching}
\end{equation}
where $g_{J\mathcal N}(q^2)$ follows from matching the up and down quark pseudoscalar densities onto nucleons~\cite{DelNobile:2021wmp}.

The corresponding nonrelativistic amplitude contains
\begin{equation}
\mathcal M_{F\mathcal N}\propto \frac{(\bm S_F\cdot\bm q)(\bm S_{\mathcal N}\cdot\bm q)}{q^2+m_J^2},
\label{eq:O6-amplitude}
\end{equation}
and therefore maps onto the nonrelativistic operator
\begin{equation}
\mathcal O_6=\left(\bm S_F\cdot \bm q\right)\left(\bm S_{\mathcal N}\cdot\bm q\right).
\end{equation}
The scattering probability consequently contains the characteristic factor 
\begin{equation}
\frac{q^4}{(q^2+m_J^2)^2},
\label{eq:q4-factor}
\end{equation}
modulated by the longitudinal spin response of the target nucleus.

The differential event rate for dark matter (DM)-nucleus scattering can be written as~\cite{Freese:2012xd}
\begin{equation}
\frac{dR}{dE_R}
=
{\cal E}\, N_T\, \mathcal{A}(E_R)\, \frac{\rho_{\rm DM}}{M_{\rm F}}
\int_{v_{\min}(E_R)}^{v_{\max}}
d^3v\,v\,f(v)\,
\frac{d\sigma}{dE_R},
\label{eq:event_rate}
\end{equation}
where \(N_T\) denotes the number of target nuclei per ton of xenon target, \(\mathcal E=2.84~\mathrm{ton\times yr}\) is the total exposure of the experiment, $\mathcal{A}(E_R)$ the recoil energy dependent efficiency adopted from \cite{LZ:2026highrecoil}, \(\rho_{\rm DM}=0.3~{\rm GeV/cm^3}\) the local DM energy density, and \(M_{\rm F}\) is the DM
particle mass. In our analysis, we consider a non-relativistic galactic
WIMP with a Maxwell-Boltzmann velocity distribution. In the laboratory
frame, the normalized speed distribution is given by
\begin{equation}
f(v)
=
\frac{v}{\sqrt{\pi}v_0 v_{\rm lab}}
\exp\left[-\frac{v^2+v_{\rm lab}^2}{v_0^2}\right]
\left[
\exp\left(\frac{2vv_{\rm lab}}{v_0^2}\right)
-
\exp\left(-\frac{2vv_{\rm lab}}{v_0^2}\right)
\right],
\label{eq:maxwell_distribution}
\end{equation}
where $v_0 = 220~\rm km/s$  and \(v_{\rm lab}\) is the velocity of the Earth with respect to the galactic rest frame.

For the pseudoscalar interaction considered here, the
 DM-nucleus differential scattering cross section is given by~\cite{DelNobile:2021wmp}
\begin{equation}
\frac{d\sigma}{dE_R}
=
\frac{g_F^2 g_u^2}{8v^2 m_F^2}
\frac{(g_A^3)^2 m_A}{(2I_A+1)(m_u+m_d)^2}
\frac{q^4 m_\pi^4}
{(q^2+m_J^2)^2(q^2+m_\pi^2)^2}
\left[
W_{\Sigma''}^{pp}
-2W_{\Sigma''}^{pn}
+W_{\Sigma''}^{nn}
\right].
\label{eq:cross_section}
\end{equation}
where \(q=\sqrt{2m_A E_R}\) is the momentum transfer, \(m_A\) is the target nuclear mass and $I_A$ the nuclear spin.  The pseudoscalar
density of the up quark contains an isovector component, which can induce the exchange of a virtual neutral pion. Consequently, the pion propagator introduces a factor \(1/(q^2+m_\pi^2)\) in the scattering amplitude. The explicit $q^4$ dependence in the differential cross section ensures that the rate vanishes in the zero momentum transfer limit. For $q\gtrsim m_\pi$, however, the momentum dependence of the pion propagator partially compensates the explicit
$q^4$ enhancement. As a result, the recoil spectrum approaches a saturated behaviour before eventually decreasing due to the suppression from the DM halo integral and the nuclear response functions. Here, $m_\pi$ denotes the mass of the virtual pion, while $g_A^3$ is the
isovector axial coupling, given by
\begin{equation}
g_A^3 = \Delta_u^{(p)}-\Delta_d^{(p)} ,
\end{equation}
with $\Delta_u^{(p)}=0.777$ and $\Delta_d^{(p)}=-0.438$. The quantity \(W_{\Sigma''}\) encodes the nuclear response associated with the longitudinal spin contribution and is obtained following Ref.~\cite{DelNobile:2021wmp}.

The velocity integration extends from the minimum speed required to produce a recoil of energy $E_R$ to the maximum DM speed in the laboratory frame,
\begin{equation}
v_{\min}(E_R)=\sqrt{\frac{m_AE_R}{2\mu_{FA}^2}},\qquad v_{\max}=v_{\rm esc}+v_{\rm lab},
\label{eq:velocity-limits}
\end{equation}
where $\mu_{FA}=M_Fm_A/(M_F+m_A)$ is the DM-nucleus reduced mass. We take $v_{\rm esc}=544~{\rm km\,s^{-1}}$ and $v_{\rm lab}=232~{\rm km\,s^{-1}}$.

\begin{figure}[t!]
\centering
\includegraphics[width=0.65\linewidth]{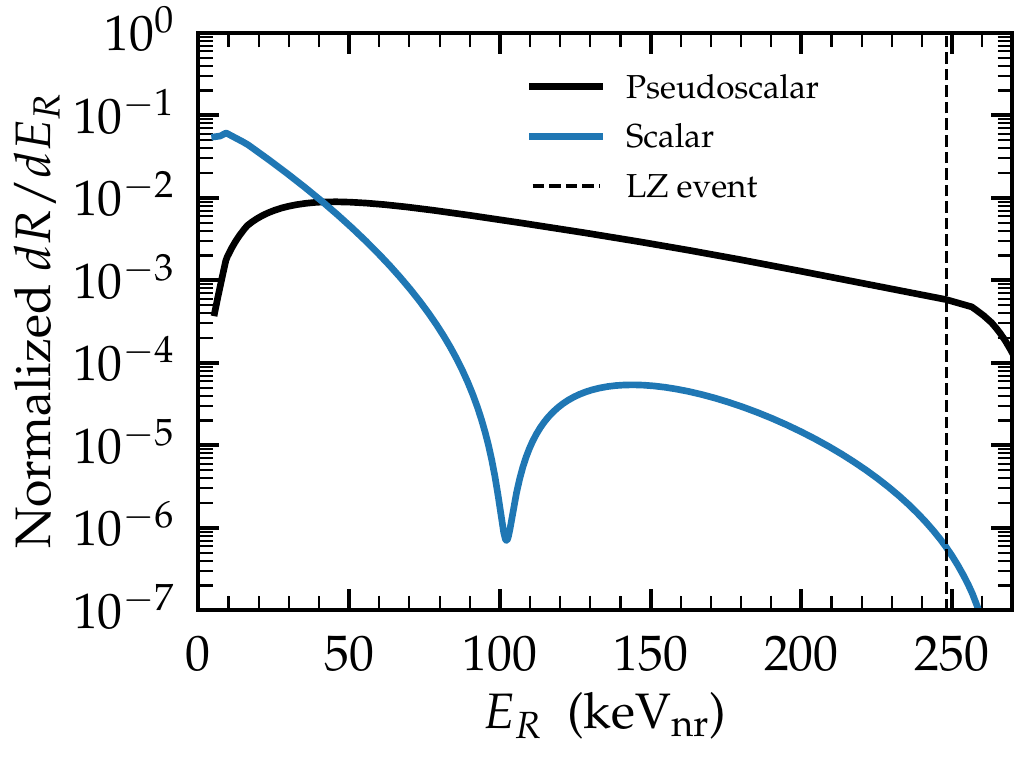}
\caption{Normalized differential recoil spectrum for dark matter scattering on xenon, assuming the $2.84~\mathrm{ton\times yr}$ exposure of the LZ dataset and a mediator coupling exclusively to up quarks. The benchmark parameters are $M_F=400~\mathrm{GeV}$, $m_J=1~\mathrm{GeV}$, and $\sqrt{g_F g_u}=0.01$. The vertical dashed line indicates the recoil energy of the LZ candidate event, $E_R=248~\mathrm{keV_{nr}}$.}
\label{fig:differential-rate}
\end{figure}

For the LZ event at $E_R\simeq 248~\mathrm{keV_{nr}}$, the characteristic momentum transfer is
\begin{equation}
q_{\rm LZ}\simeq \sqrt{2m_{A}E_R}\simeq 250~{\rm MeV}.
\label{eq:q-LZ}
\end{equation}
Fig.~\ref{fig:differential-rate} illustrates the resulting differential recoil spectra for representative pure up quark couplings of scalar and pseudoscalar mediators. As shown in the figure, scalar mediated interactions predominantly yield recoils at lower energies. The absence of a significant excess in this region therefore disfavors elastic DM-nucleus scattering through a scalar mediator as an explanation of the LZ candidate event. A similar qualitative behavior is obtained for elastic scattering mediated by vector and axial-vector interactions.

The pseudoscalar-mediated case, however, is qualitatively different. The momentum dependence of the pseudoscalar interaction shifts a substantial fraction of the signal towards the higher-recoil region probed by LZ. For mediator masses in the sub-GeV to GeV range, the interaction consequently interpolates between the light-mediator and contact-interaction regimes over the range of momentum transfers relevant to the observed event. As demonstrated by the LZ collaboration, this momentum dependent enhancement of the high recoil contribution allows pseudoscalar mediated elastic DM-nucleus scattering to provide a viable statistical description of the candidate event without simultaneously producing an excessive number of events in the low-energy recoil region.

To further investigate this distinctive feature of pseudoscalar mediated elastic scattering, we divide the recoil-energy range into bins whose widths are chosen to be twice the detector resolution evaluated at the corresponding bin centers, following the prescription adopted in Ref.~\cite{Chattaraj:2026fxn}. In particular, the bin centered at $248~\mathrm{keV_{nr}}$ has a width of $46~\mathrm{keV_{nr}}$ and therefore extends from $225$ to $271~\mathrm{keV_{nr}}$. This bin defines our signal region, as it contains the candidate event reported by the LZ collaboration. We extend the same binning prescription towards lower recoil energies, truncating the lowest bin at the analysis threshold, $T_{\mathcal{N}}^{\rm reco}=5.4~\mathrm{keV_{nr}}$.

For the analysis, we divide the full recoil-energy window, $E_R\in[5.4,271]~\mathrm{keV_{nr}}$, into two regions. Region I corresponds to the lower-recoil interval, $E_R\in[5.4,225]~\mathrm{keV_{nr}}$, while Region II corresponds to the signal region, $E_R\in[225,271]~\mathrm{keV_{nr}}$. For each point in parameter space, we fix the overall signal normalization by requiring one expected event in Region II,
\begin{equation}
R_{\rm II}=\int_{225~\mathrm{keV_{nr}}}^{271~\mathrm{keV_{nr}}}dE_R\,\frac{dR}{dE_R}=1.
\label{eq:RII}
\end{equation}

We adopt a deliberately restrictive counting criterion for the lower-recoil region. LZ observes 1710 events in the full WIMP-search region, with a total post-fit expectation of $1713\pm39$ events~\cite{LZ:2026highrecoil}. We require
\begin{equation}
R_{\rm I}=\int_{5.4~\mathrm{keV_{nr}}}^{225~\mathrm{keV_{nr}}}dE_R\,\frac{dR}{dE_R}<39.
\label{eq:RI}
\end{equation}
The allowed signal contribution is therefore below the quoted uncertainty on the total event yield. Together with the normalization condition $R_{\rm II}=1$, this criterion selects spectra that account for the high-recoil candidate without producing a sizeable excess in the background dominated lower-recoil region\footnote{Since the LZ Collaboration has not yet provided the detailed binning information or the detector response and resolution as a function of nuclear-recoil energy, a fully consistent likelihood analysis cannot be performed within the scope of the present work. We therefore adopt the conservative event counting criteria $
R_{\rm II}=1, \, R_{\rm I}<39$, to identify the region of parameter space that can accommodate the high-recoil candidate while remaining consistent with the observed event yield in the lower-recoil region. A dedicated statistical analysis incorporating the complete detector response, energy resolution, binning, and likelihood information would be required to derive a quantitatively robust constraint on the model parameters and can only be performed once these details are made public by the LZ Collaboration.}.

\begin{figure*}[ht!]
\centering
\includegraphics[width=0.47\textwidth]{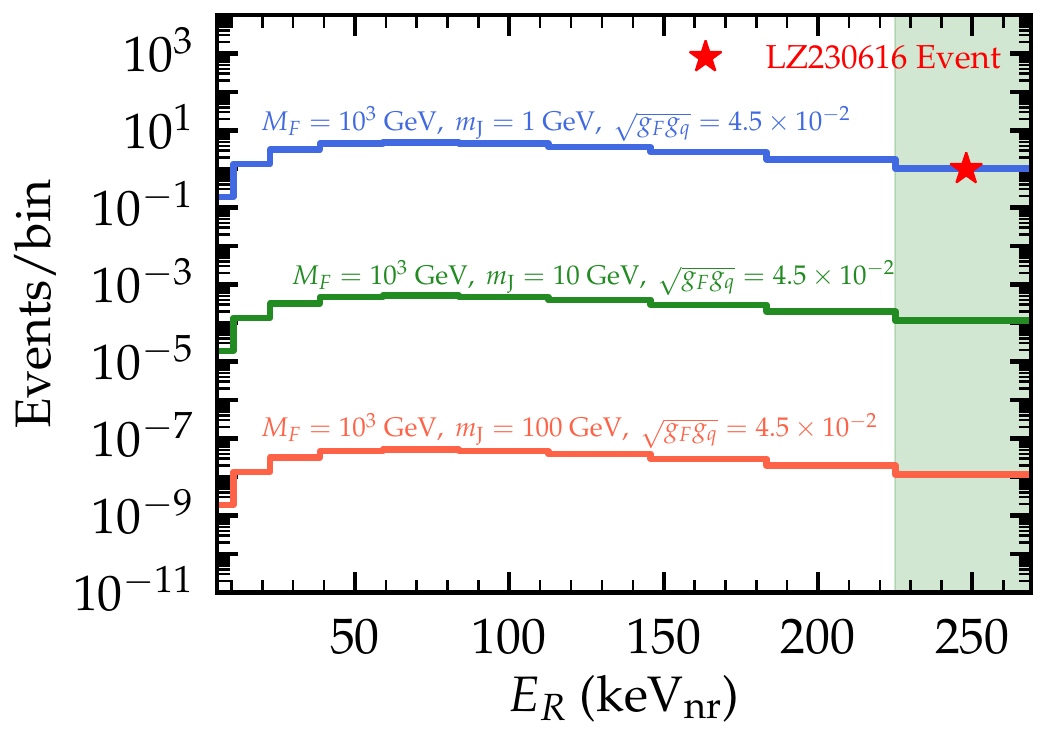}
\includegraphics[width=0.47\textwidth]{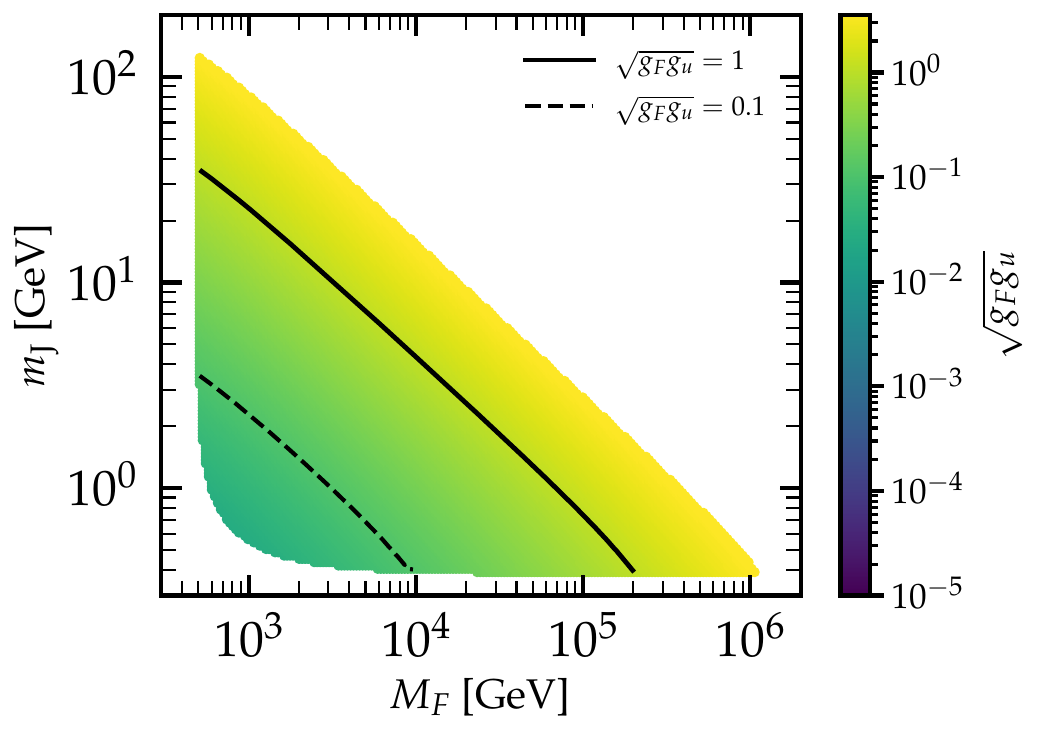}
\caption{\textit{Left panel:} Predicted nuclear recoil spectra for elastic DM-nucleus scattering mediated by $J$, for representative values of the dark matter mass $M_F$, mediator mass $m_J$ and coupling combination. The red star denotes the LZ candidate event. \textit{Right panel:} Regions in the $(M_F,m_J)$ plane satisfying $R_{\rm II}=1$ and $R_{\rm I}<39$, with the colour scale indicating the coupling combination required by the normalization. For both panels we have assumed a pure up quark coupling and have taken $g_d=0$. }
\label{fig:model_independent}
\end{figure*}

The resulting spectra and allowed regions is shown in Fig.~\ref{fig:model_independent}. The left panel illustrates how the mediator mass controls the recoil dependence, while the right panel shows the parameter space selected by the aforementioned counting requirements. For $m_J^2\ll q^2$, the factor $q^4/(q^2+m_J^2)^2$ approaches a constant and the suppression of low-energy recoils is weakened. Increasing $m_J$ relative to the momentum transfers probed by LZ restores the characteristic $q^4$ weighting and shifts a larger fraction of the signal into Region II. The upper extent of the region shown in the right panel is instead set by the coupling required to maintain $R_{\rm II}=1$, restricting the displayed coupling combination to values below unity. The different up and down quark assignments modify the longitudinal xenon response and shift the allowed region, while preserving the preference for a mediator substantially lighter than the dark matter particle. For Fig..~\ref{fig:model_independent} we have assumed a purely up quark coupling and have taken Majoron-down-quark coupling $g_d = 0$.

These properties point naturally towards a massive pseudo-Nambu-Goldstone mediator. The same spin and momentum dependence also changes the hierarchy of solar capture targets, as we discuss next.

\subsection{Solar capture}

For conventional spin independent scattering of heavy dark matter, solar capture is efficiently driven by heavy elements~\cite{Nussinov:2009ft,Pospelov:2026solar,DiMauroShaikh:2026solar}. In particular, the element by element analysis of Ref.~\cite{Blennow:2015hzp} shows that iron provides the leading contribution over much of the parameter space relevant to endothermic dark matter. Solar iron is overwhelmingly composed of ${}^{56}{\rm Fe}$, which has zero nuclear spin. The pseudoscalar interaction considered here therefore leads to a very different hierarchy of solar targets. The operator $\mathcal O_6$ probes the longitudinal nuclear spin response, associated with the component of the nuclear spin parallel to the momentum transfer, and vanishes for nuclei with $I_A=0$~\cite{Fitzpatrick:2012ix,Anand:2013yka,CatenaSchwabe:2015solar}.

Consequently, the response to $\mathcal O_6$ differs between xenon detectors and solar matter. Almost half of natural xenon is composed of the isotopes ${}^{129}{\rm Xe}$ and ${}^{131}{\rm Xe}$, which have $I_A=1/2$ and $3/2$, respectively. By contrast, the average solar mass fractions of ${}^{1}{\rm H}$, ${}^{4}{\rm He}$, ${}^{16}{\rm O}$, ${}^{14}{\rm N}$ and ${}^{56}{\rm Fe}$ are approximately $0.684$, $0.298$, $8.5\times10^{-3}$, $1.6\times10^{-3}$ and $1.4\times10^{-3}$, respectively~\cite{CatenaSchwabe:2015solar}, and ${}^{4}{\rm He}$, ${}^{16}{\rm O}$ and ${}^{56}{\rm Fe}$ all have $I_A=0$. The dominant iron capture channel is therefore absent for $\mathcal O_6$, leaving hydrogen and the much less abundant nuclei with nonzero spin as the relevant solar targets.

Hydrogen is the most abundant solar target with nonzero nuclear spin, but its ability to capture heavy dark matter is limited by both kinematics and momentum suppression. A particle arriving at the Sun with asymptotic velocity $u$ reaches a velocity $w$ at radius $r$. In a single elastic collision with a nucleus $A$, capture is possible only if the maximal fractional energy loss exceeds the energy that must be dissipated to become gravitationally bound~\cite{Gould:1987ir,Gould:1987ju}:
\begin{equation}
w^2=u^2+v_{\rm esc}^2(r),\qquad
\beta_A=\frac{4M_{\rm F}\,m_A}{(M_{\rm F}+m_A)^2},\qquad
u<u_{\rm max}^A(r)\equiv v_{\rm esc}(r)\sqrt{\frac{\beta_A}{1-\beta_A}}.
\label{eq:solar capture-kinematics}
\end{equation}
For the few-hundred-GeV to TeV dark matter masses relevant to the high-recoil event, $\beta_H$ lies between a few $10^{-3}$ and $10^{-2}$. Even at the solar centre, where $v_{\rm esc}(0)\simeq1.4\times10^3~{\rm km\,s^{-1}}$, this restricts hydrogen capture to approximately $u\lesssim90$-$170~{\rm km\,s^{-1}}$. Only the low velocity part of the incident population can therefore be captured on hydrogen, with an even stronger restriction away from the solar centre.

For hydrogen, the momentum dependence of pseudoscalar exchange produces an additional suppression. Since the reduced mass of the DM-proton system is close to the proton mass, the largest momentum transfer attainable in the solar core is $q_{\rm max}^{H}\simeq2\mu_Hw\lesssim9~{\rm MeV}$, well below the characteristic value $q_{\rm LZ}\simeq250~{\rm MeV}$. The suppression from the derivative interaction and the mediator propagator alone can be displayed through
\begin{equation}
\mathcal S_H(m_J)\equiv
\frac{\left(q_{\rm max}^{H}\right)^4}{\left[\left(q_{\rm max}^{H}\right)^2+m_J^2\right]^2}
\frac{\left(q_{\rm LZ}^2+m_J^2\right)^2}{q_{\rm LZ}^4}
\sim10^{-6}\text{-}10^{-5},
\label{eq:hydrogen-suppression}
\end{equation}
where the estimate applies to mediator masses of a few tenths of a GeV, as selected by the high-recoil spectrum. This ratio is independent of the overall coupling normalization. Since $q_{\rm max}^{H}$ is an upper bound, Eq.~\eqref{eq:hydrogen-suppression} provides a conservative estimate of this effect. 

Residual capture is controlled by the much less abundant $I_A\neq0$ isotopes. Among them, ${}^{14}{\rm N}$ is particularly relevant because its larger mass improves the kinematic matching to heavy dark matter: momentum transfers of order $100~{\rm MeV}$ are possible and particles with asymptotic velocities of several hundred ${\rm km\,s^{-1}}$ can be captured. Writing $X_A$ for the solar mass fraction of isotope $A$, its number density relative to hydrogen is nevertheless suppressed by
\begin{equation}
\frac{n_{{}^{14}{\rm N}}}{n_H}\simeq\frac{X_{{}^{14}{\rm N}}}{14X_H}\simeq1.7\times10^{-4},
\label{eq:nitrogen-abundance}
\end{equation}
Other contributing isotopes, including ${}^{3}{\rm He}$, ${}^{13}{\rm C}$, ${}^{15}{\rm N}$, ${}^{17}{\rm O}$, ${}^{23}{\rm Na}$ and ${}^{27}{\rm Al}$, are similarly scarce. Determining their relative contributions requires integrating the corresponding longitudinal spin responses over the solar density and velocity profiles. Complete non-relativistic effective field theory (NREFT) calculations identify hydrogen and ${}^{14}{\rm N}$ as the principal targets for $\mathcal O_6$ capture at large dark matter masses~\cite{CatenaSchwabe:2015solar}. The resulting suppression is supported by the dedicated solar capture analysis of Ref.~\cite{Bose:2026solar}. That study includes the full set of solar isotopes and confronts the resulting neutrino fluxes with IceCube and Super-Kamiokande data. Neither the isoscalar nor the isovector pseudoscalar-pseudoscalar interaction $\mathcal L_4$, which reduces to $\mathcal O_6$, is excluded in the LZ favoured region, even for a unit branching fraction into the maximally constraining $\nu\bar\nu$ final state. 

We conclude that tree level Majoron exchange strongly suppresses solar capture through three simultaneous effects: the principal heavy element targets have zero nuclear spin, hydrogen suffers from poor kinematic matching and small momentum transfer, and the remaining nuclei with nonzero spin have low solar abundances. Loop induced scalar operators could in principle open spin independent capture channels on zero-spin nuclei~\cite{Bell:2026pseudoscalarSI}, but their size is model dependent and its precise calculation is beyond the scope of this work.
We now embed this pseudoscalar interaction in a radiative neutrino mass model, identifying the mediator with the Majoron associated with spontaneous lepton number breaking.

\section{A scotogenic Majoron realization}
\label{sec:scoto}

\subsection{Field content and symmetry breaking}

We extend the scotogenic model by a complex scalar singlet $\Phi$ and one up-type vector-like quark $U$. The model contains an exact dark $\mathbb Z_2$ and a global $U(1)_X$ symmetries i.e. $U(1)_X \equiv U(1)_L$. In the lepton and scotogenic sectors, $U(1)_X$ acts as lepton number. It is spontaneously broken by $\langle\Phi\rangle$, while the exact $\mathbb Z_2$ stabilizes the lightest odd state. We take this state to be the lightest singlet Majorana fermion $N_1$. The charge assignment is given in Table~\ref{tab:charges}.

\begin{table}[t]
\centering
\renewcommand{\arraystretch}{1.08}
\begin{tabular}{lccc}
\toprule
\hspace{0.1cm}Field \hspace{0.1cm}&\hspace{0.1cm} $SU(3)_c\times SU(2)_L\times U(1)_Y$ \hspace{0.1cm}&\hspace{0.1cm} $U(1)_X \equiv U(1)_L$ \hspace{0.1cm}&\hspace{0.1cm} $\mathbb Z_2$\hspace{0.1cm} \\
\midrule
\multicolumn{4}{l}{\textbf{Standard Model sector}} \\
\midrule
$Q_L$ & $(\mathbf 3,\mathbf 2,1/6)$ & $0$ & $+$ \\
$(u_R,c_R,t_R)$ & $(\mathbf 3,\mathbf 1,2/3)$ & $(2,0,0)$ & $+$ \\
$d_R$ & $(\mathbf 3,\mathbf 1,-1/3)$ & $0$ & $+$ \\
$L$ & $(\mathbf 1,\mathbf 2,-1/2)$ & $1$ & $+$ \\
$e_R$ & $(\mathbf 1,\mathbf 1,-1)$ & $1$ & $+$ \\
$H$ & $(\mathbf 1,\mathbf 2,1/2)$ & $0$ & $+$ \\
\midrule
\multicolumn{4}{l}{\textbf{Scotogenic sector}} \\
\midrule
$\eta$ & $(\mathbf 1,\mathbf 2,1/2)$ & $0$ & $-$ \\
$N_i$ & $(\mathbf 1,\mathbf 1,0)$ & $1$ & $-$ \\
$\Phi$ & $(\mathbf 1,\mathbf 1,0)$ & $-2$ & $+$ \\
\midrule
\multicolumn{4}{l}{\textbf{Vector-like quark sector}} \\
\midrule
$U_L$ & $(\mathbf 3,\mathbf 1,2/3)$ & $0$ & $+$ \\
$U_R$ & $(\mathbf 3,\mathbf 1,2/3)$ & $0$ & $+$ \\
\bottomrule
\end{tabular}
\caption{Field content and charge assignment. The right handed up-type quarks carry family dependent $U(1)_X$ charges, ultimately leading to a Majoron-up-quark interaction.}
\label{tab:charges}
\end{table}

The scotogenic fermion interactions have their usual form, with the singlet Majorana masses generated by spontaneous lepton number breaking:
\begin{equation}
-\mathcal L_{\rm sc}\supset h_{\alpha i}\,\overline{L_\alpha}\widetilde\eta N_i+\frac{1}{2}(Y_N)_{ij}\Phi\,\overline{N_i^c}N_j+(Y_e)_{\alpha\beta}\overline{L_\alpha}H e_{R\beta}+{\rm h.c.}
\end{equation}
Without loss of generality, we work in a basis where $Y_e$ and the symmetric matrix $Y_N$ are both diagonal and real. In the quark sector, the charge of $u_R$ forbids a direct Higgs Yukawa coupling to it. The allowed terms are
\begin{align}
-\mathcal L_q\supset {}&
\sum_{i=1}^3\overline{Q_{Li}}\widetilde H\bigl(y_{ic}c_R+y_{it}t_R+y_{iU}U_R\bigr)
+\overline{U_L}\left[\sum_{f=c,t,U}\vec M f_R+y_U\Phi u_R\right] \nonumber\\
&+\sum_{i,j=1}^3(Y_d)_{ij}\overline{Q_{Li}}H d_{Rj}+{\rm h.c.}
\label{eq:yukawa}
\end{align}
Here $\vec M\equiv(M_c,M_t,M_U)^T$, with components ordered as $(c,t,U)$. We can without loss of generality go to the basis in which the Higgs Yukawa couplings take the form $y_{uU}\overline{Q_{L1}}\widetilde H U_R+y_c\overline{Q_{L2}}\widetilde H c_R+y_t\overline{Q_{L3}}\widetilde H t_R$. In this basis, $\vec M$ and $Y_d$ remain general. We then take the alignment Ansatz $\vec M=M_U(0,0,1)^T$, so that the CKM mixing comes from the down sector.

The $U(1)_X$ invariant scalar potential contains
\begin{align}
V_{\rm inv}={}&-m_H^2H^\dagger H+m_\eta^2\eta^\dagger\eta-m_\Phi^2\Phi^\dagger\Phi
+\frac{\lambda_H}{2}(H^\dagger H)^2+\frac{\lambda_\eta}{2}(\eta^\dagger\eta)^2+\frac{\lambda_\Phi}{2}(\Phi^\dagger\Phi)^2 \nonumber\\
&+\lambda_3(H^\dagger H)(\eta^\dagger\eta)+\lambda_4(H^\dagger\eta)(\eta^\dagger H)
+\left[\frac{\lambda_5}{2}(H^\dagger\eta)^2+{\rm h.c.}\right] \nonumber\\
&+\lambda_{H\Phi}(H^\dagger H)(\Phi^\dagger\Phi)
+\lambda_{\eta\Phi}(\eta^\dagger\eta)(\Phi^\dagger\Phi).
\label{eq:scalar-potential}
\end{align}
We introduce explicit $U(1)_X$ soft breaking through
\begin{equation}
-\mathcal L_{\rm soft}\supset M_U^0\,\overline{U_L}u_R+{\rm h.c.},
\qquad
V_{\rm soft}\supset-\frac{\mu_\Phi^2}{2}\bigl(\Phi^2+\Phi^{\dagger 2}\bigr).
\end{equation}

We parametrize the symmetry breaking field as
\begin{equation}
\Phi=\frac{v_\phi+\rho}{\sqrt{2}}e^{iJ/v_\phi}.
\label{eq:phi}
\end{equation}
Without the soft breaking terms, $J$ is the Goldstone boson of $U(1)_X$ at the classical level. The scalar soft term gives it a mass, and $M_U^0$ generates further symmetry breaking contributions radiatively. We treat the renormalized $m_J$ as a phenomenological input.

\subsection{Majorana dark matter and the up quark portal}

After spontaneous symmetry breaking, the singlet fermion mass matrix is
\begin{equation}
M_N=\frac{v_\phi}{\sqrt{2}}Y_N=\operatorname{diag}(M_1,M_2,M_3).
\end{equation}
We assume $M_1<m_\eta,M_2,M_3$, so that $N_1$ is the dark matter candidate. The same Yukawa interaction determines the Majoron coupling to the singlet fermions,
\begin{equation}
\mathcal L_J\supset-\frac{iJ}{2v_\phi}\sum_i M_i\,\overline N_i\gamma_5N_i.
\label{eq:JNN}
\end{equation}

In the up quark sector, we consider the limit of small left and right handed mixing angles, $s_L\equiv\sin\theta_L\ll1$ and $s_R\equiv\sin\theta_R\ll1$. We denote the Higgs-induced mass entries by $m_u^{(0)}\equiv y_{uU}v/\sqrt{2}$, $m_c^{(0)}\equiv y_cv/\sqrt{2}$ and $m_t^{(0)}\equiv y_tv/\sqrt{2}$, and define
\begin{equation}
\mu_{\rm eff}\equiv\frac{y_Uv_\phi}{\sqrt{2}}+M_U^0.
\end{equation}
Small mixing corresponds to $M_U\gg m_u^{(0)},|\mu_{\rm eff}|$. With the alignment Ansatz specified above, the mass matrix in the basis $(u_L,c_L,t_L,U_L)$ for the left handed fields and $(u_R,c_R,t_R,U_R)$ for the right handed fields is
\begin{equation}
\mathcal M_{uU}=
\begin{pmatrix}
0&0&0&m_u^{(0)}\\
0&m_c^{(0)}&0&0\\
0&0&m_t^{(0)}&0\\
\mu_{\rm eff}&0&0&M_U
\end{pmatrix}.
\label{eq:up-mass-matrix}
\end{equation}
Charm and top do not mix with $U$, leaving the type-I Dirac seesaw block
\begin{equation}
\mathcal M_{uU}^{(2)}=
\begin{pmatrix}
0&m_u^{(0)}\\
\mu_{\rm eff}&M_U
\end{pmatrix}.
\label{eq:up-seesaw-block}
\end{equation}
To leading order in the mixing angles,
\begin{align}
m_u&\simeq\frac{m_u^{(0)}\mu_{\rm eff}}{M_U},
&
m_{U,\mathrm{phys}}&\simeq M_U\left[1+\frac{(m_u^{(0)})^2+\mu_{\rm eff}^2}{2M_U^2}\right],
\nonumber\\
s_L&\simeq\frac{m_u^{(0)}}{M_U},
&
s_R&\simeq\frac{\mu_{\rm eff}}{M_U}\simeq\frac{m_u}{m_u^{(0)}}.
\label{eq:heavy-U-mass}
\end{align}
For a chosen $m_u^{(0)}$, the measured up quark mass determines $\mu_{\rm eff}\simeq M_Um_u/m_u^{(0)}$ and hence
\begin{equation}
M_U^0\simeq\frac{M_Um_u}{m_u^{(0)}}-\frac{y_Uv_\phi}{\sqrt{2}}.
\label{eq:up-soft-mass}
\end{equation}
The right handed rotation leaves the Standard Model gauge currents unchanged because $u_R$ and $U_R$ have identical gauge charges, while corrections to the left handed electroweak currents are of order $s_L^2$.

Only the $\Phi$ dependent part of $\mu_{\rm eff}$ couples to the Majoron. At leading order in the mixing angles, the interaction $-\mathcal L_J\supset i(y_U/\sqrt{2})J\overline U_Lu_R+{\rm h.c.}$ gives
\begin{equation}
\mathcal L_J\supset-i\frac{y_U}{\sqrt{2}}J\left(s_L\,\overline u\gamma_5u+s_R\,\overline U\gamma_5U\right)+\mathcal L_J^{\mathrm{off\text{-}diag}}.
\label{eq:J-quark-couplings}
\end{equation}
The off-diagonal interaction permits $U\to uJ$, while the direct Majoron coupling to the physical up quark is
\begin{equation}
g_{Juu}\simeq\frac{y_Um_u^{(0)}}{\sqrt{2}M_U}.
\label{eq:Juu}
\end{equation}
Integrating out $U$ also generates a gluonic operator. Using $m_u\simeq s_Ls_RM_U$, its contribution relative to the direct up quark contribution scales as
\begin{equation}
\frac{g_{JUU}/m_{U,\mathrm{phys}}}{g_{Juu}/m_u}\simeq s_R^2\ll1,
\label{eq:Jgg-suppression}
\end{equation}
where $g_{JUU}\simeq y_Us_R/\sqrt{2}$. We therefore neglect the heavy-quark-induced gluonic contribution in the small mixing regime.

We restrict the Higgs-induced entry to $m_u\ll m_u^{(0)}<m_c^{(0)}$, with all running quantities evaluated at a common renormalization scale. This ensures
\begin{equation}
s_R\simeq\frac{m_u}{m_u^{(0)}}\ll1,\qquad
s_L\simeq\frac{m_u^{(0)}}{M_U}<\frac{m_c^{(0)}}{M_U}\ll1.
\end{equation}

For concreteness, we fix $M_U=2~\mathrm{TeV}$, which implies $m_{U,\mathrm{phys}}>2~\mathrm{TeV}$. For orientation, ATLAS excludes light flavour vector-like quarks below $1.53~\mathrm{TeV}$ when ${\cal B}(Q\to Wq)=1$~\cite{ATLAS:2024VLQlight}. Since the $U\to uJ$ decay considered here leads to a different collider signature, we use this result as an indication of the relevant mass scale constraint.

In summary, the resulting low energy quark and DM Majoron portal is given by
\begin{equation}
\boxed{\mathcal L_{\rm portal}\simeq-\frac{iM_1}{2v_\phi}J\overline N_1\gamma_5N_1-i\frac{y_Um_u^{(0)}}{\sqrt{2}M_U}J\overline u\gamma_5u.}
\label{eq:minimal-portal}
\end{equation}
We fix $M_U=2~\mathrm{TeV}$ and take $m_u^{(0)}$ and $y_U$ as free scan parameters in the ranges
\begin{equation}
m_u= 2.2 \mathrm{MeV} \ll m_u^{(0)} \lesssim  m_c^{(0)}\simeq600~\mathrm{MeV},
\qquad
0.1\leq y_U\leq1.
\end{equation}
For each parameter point, $\mu_{\rm eff}$ and $M_U^0$ are fixed by the measured up quark mass through Eq.~\eqref{eq:up-soft-mass}, and we retain points satisfying $s_L,s_R\ll1$.

\subsection{Radiative neutrino masses}

The neutrino mass mechanism is the standard scotogenic one. Writing $\eta^0=(\eta_R+i\eta_I)/\sqrt{2}$, the neutral inert-scalar masses are
\begin{equation}
m_{\eta_R}^2=m_\eta^2+\frac{1}{2}(\lambda_3+\lambda_4+\lambda_5)v^2,\qquad
m_{\eta_I}^2=m_\eta^2+\frac{1}{2}(\lambda_3+\lambda_4-\lambda_5)v^2.
\end{equation}
The one loop neutrino mass matrix is
\begin{align}
(m_\nu)_{\alpha\beta}=\sum_i\frac{h_{\alpha i}h_{\beta i}M_i}{32\pi^2}\bigg[&
\frac{m_{\eta_R}^2}{m_{\eta_R}^2-M_i^2}\ln\left(\frac{m_{\eta_R}^2}{M_i^2}\right)
-\frac{m_{\eta_I}^2}{m_{\eta_I}^2-M_i^2}\ln\left(\frac{m_{\eta_I}^2}{M_i^2}\right)\bigg].
\label{eq:neutrino mass}
\end{align}
The parameters controlling neutrino masses coexist with the Majoron portal without modifying the elastic nature of the $N_1$-nucleus interaction.

\section{Phenomenology}
\subsection{The LZ high-recoil event}

\begin{figure}
    \centering
    \includegraphics[width=0.47\textwidth]{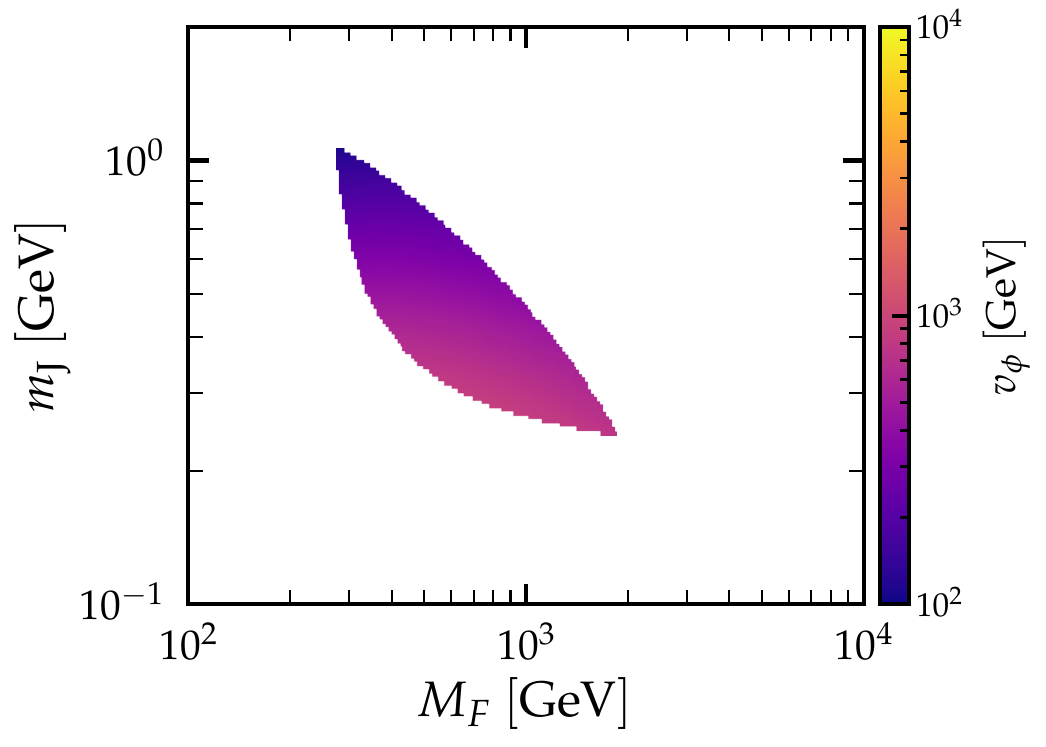}
    \hspace{0.02\textwidth}
    \includegraphics[width=0.47\textwidth]{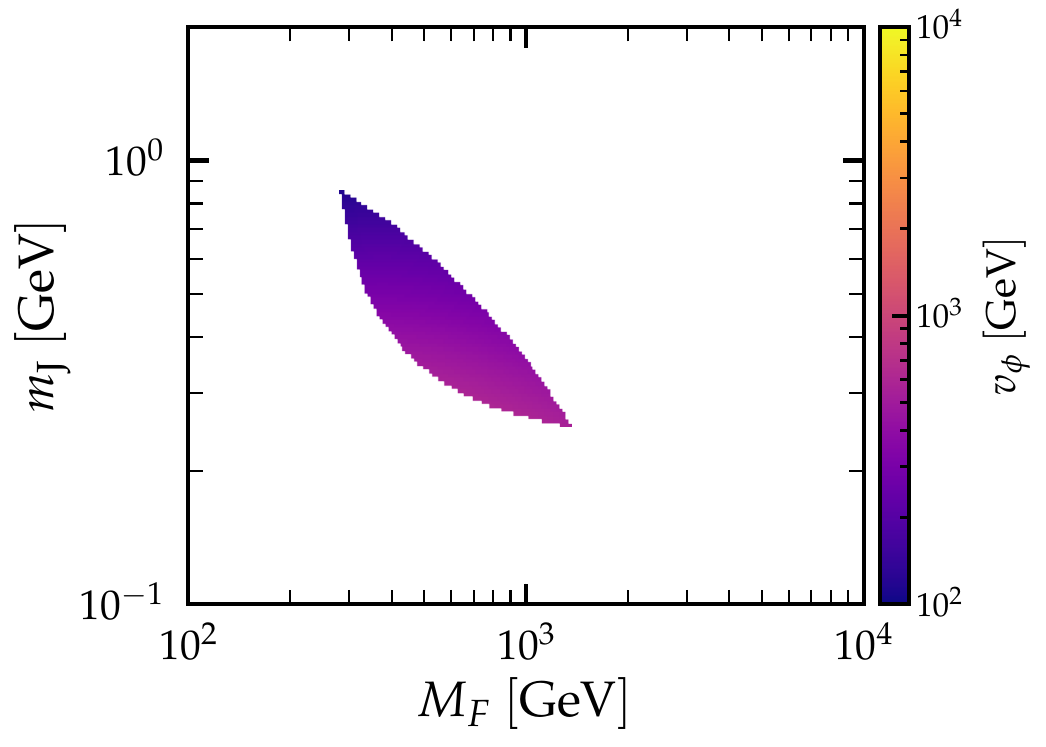}
    \caption{Allowed parameter space in the $(M_1,m_J)$ plane for $y_Um_u^{(0)}=600~\mathrm{MeV}$ (left) and $y_Um_u^{(0)}=400~\mathrm{MeV}$ (right), obtained by requiring one expected event in the LZ signal region, $R_{\rm II}=1$, while restricting the predicted number of events in the lower recoil region to $R_{\rm I}<39$. The colour scale indicates the corresponding value of the Majoron vev $v_\phi$ in the allowed parameter space.}
    \label{fig:3}
\end{figure}

The low-energy portal of Eq.~\eqref{eq:minimal-portal} provides a concrete realization of the model independent interaction introduced in Eq.~\eqref{eq:general-portal}. Identifying $F=N_1$, the relevant couplings are
\begin{equation}
    g_F=\frac{M_1}{2v_\phi},\,\qquad g_u=\frac{y_Um_u^{(0)}}{\sqrt{2}M_U},\qquad g_d=0.
    \label{eq:model-couplings}
\end{equation}
The model therefore realizes directly the pure up quark scenario studied in Sec.~\ref{subsec:elastic-high-recoil}. In particular, the scattering is governed by the non-relativistic operator $\mathcal O_6$, and the recoil spectra and nuclear responses derived there apply directly.

For fixed $M_U=2~\mathrm{TeV}$, the direct detection signal depends on $M_1$, $m_J$, $v_\phi$ and the combination $y_Um_u^{(0)}$. We apply the same procedure as in the model independent analysis. For every $(M_1,m_J)$ point, the overall normalization is fixed by requiring one expected event in the high-recoil signal region, $R_{\rm II}=1$, as defined in Eq.~\eqref{eq:RII}, and we retain only points satisfying the lower-recoil requirement $R_{\rm I}<39$ of Eq.~\eqref{eq:RI}.

The resulting parameter space is shown in Fig.~\ref{fig:3} for $y_Um_u^{(0)}=600~\mathrm{MeV}$ and $400~\mathrm{MeV}$ in the left and right panels, respectively. The colour scale shows the value of $v_\phi$ required to reproduce the LZ normalization. As expected from the model independent results of Fig.~\ref{fig:model_independent}, the viable region selects a light Majoron with a mass in the sub-GeV range. For fixed $(M_1,m_J)$, reducing $y_Um_u^{(0)}$ decreases the Majoron coupling to the up quark and therefore requires a smaller $v_\phi$, corresponding to a larger Majoron coupling to dark matter, in order to maintain $R_{\rm II}=1$.

The parameter region selected by LZ is also consistent with the small mixing approximation used to derive Eq.~\eqref{eq:Juu}. For any choice of $m_u^{(0)}$ and $y_U$, the physical up quark mass fixes $\mu_{\rm eff}$ and, consequently, the soft breaking mass $M_U^0$. Together with the mixing angles in Eq.~\eqref{eq:heavy-U-mass}, one finds
\begin{align}
    \mu_{\rm eff}&\simeq\frac{M_Um_u}{m_u^{(0)}}, &
    M_U^0&\simeq\frac{M_Um_u}{m_u^{(0)}}-\frac{y_Uv_\phi}{\sqrt{2}}, \nonumber\\
    s_L&\simeq\frac{m_u^{(0)}}{M_U}, &
    s_R&\simeq\frac{m_u}{m_u^{(0)}}.
\end{align}
For instance, taking $y_U=1$ and $m_u^{(0)}=600~\mathrm{MeV}$, corresponding to the left panel of Fig.~\ref{fig:3}, gives
\begin{equation}
    s_L\simeq2.0\times10^{-4},\qquad s_R\simeq3.7\times10^{-3},\qquad \mu_{\rm eff}\simeq11~\mathrm{GeV}.
\end{equation}
Both mixing angles are safely within the small mixing regime, while the measured up quark mass is reproduced by construction. Furthermore, Eq.~\eqref{eq:Jgg-suppression} implies that the heavy $U$ contribution to the effective gluonic coupling is suppressed relative to the direct up quark contribution by $s_R^2\simeq1.3\times10^{-5}$ for this choice.

The LZ signal can therefore be reproduced within the regime in which the approximations used to derive the low energy portal in Eq.~\eqref{eq:minimal-portal} remain valid. The remaining question is whether the same region of $(M_1,m_J,v_\phi)$ parameter space can simultaneously account for the observed dark matter abundance. We address this in Sec.~\ref{subsec:relic}.

\subsection{Thermal relic density}
\label{subsec:relic}
To evaluate the cosmological viability of the model, we investigate constraints arising from the dark matter relic density. The model was implemented in \texttt{SARAH}~\cite{Staub:2013tta,Staub:2015kfa} to generate a \texttt{SPheno}~\cite{Porod:2003um, Porod:2011nf} spectrum generator along with a \texttt{CalcHEP} file for \texttt{micrOMEGAs}~\cite{Belanger:2018ccd, Alguero:2023zol}, which was utilized to evaluate the thermal relic abundance of the dark matter $N_1$.

Prior to computing the relic density, we enforce theoretical consistency conditions on the scalar potential. Requiring the potential to be bounded from below yields the copositivity conditions~\cite{Kannike:2012pe}
\begin{equation}
\label{eq:copos}
\begin{gathered}
  \lambda_H > 0\,, \qquad \lambda_\eta > 0\,, \qquad \lambda_\Phi > 0\,, \\[6pt]
  \tilde\lambda_3 > -2\sqrt{\lambda_H \lambda_\eta}\,, \qquad
   \lambda_{H\Phi} > -2\sqrt{\lambda_H \lambda_\Phi}\,, \qquad
   \lambda_{\eta\Phi} > -2\sqrt{\lambda_\eta \lambda_\Phi}\,, \\[6pt]
  2\sqrt{\lambda_H \lambda_\eta \lambda_\Phi}
   + \tilde\lambda_3 \sqrt{\lambda_\Phi}
   + \lambda_{H\Phi} \sqrt{\lambda_\eta}
   + \lambda_{\eta\Phi} \sqrt{\lambda_H} \\[2pt]
   + \sqrt{
     \bigl(\tilde\lambda_3 + 2\sqrt{\lambda_H\lambda_\eta}\bigr)
     \bigl(\lambda_{H\Phi} + 2\sqrt{\lambda_H\lambda_\Phi}\bigr)
     \bigl(\lambda_{\eta\Phi} + 2\sqrt{\lambda_\eta\lambda_\Phi}\bigr)
   } > 0\,.
\end{gathered}
\end{equation}
where 
\begin{equation}
  \tilde\lambda_3 \equiv \lambda_3 + \min\left(0,\, \lambda_4 - |\lambda_5|\right)
\end{equation}
accounts for the orientation dependence of the two doublets under the $\lambda_4$ and $\lambda_5$ quartic terms. We further require the vacuum configuration $\langle H \rangle = v/\sqrt{2}$, $\langle \Phi \rangle = v_\phi/\sqrt{2}$, and $\langle \eta \rangle = 0$ to constitute a global minimum, ensuring all physical scalar masses are positive and that the CP even state $\rho$ is heavier than the SM like Higgs boson. The coupling $\lambda_H$ is fixed by matching the lighter CP even eigenstate to the observed Higgs mass, $m_h = 125.09 \pm 0.11 ~\mathrm{GeV}$~\cite{ATLAS:2023oaq}. 

All couplings are required to remain perturbative,
\begin{equation}
  |\lambda_i| \leq \sqrt{4\pi}\,, \qquad |Y_{N_k}|,\ |(y_\nu)_{\alpha k}| \leq \sqrt{4\pi}\,,
  \label{eq:pert}
\end{equation}
where $\lambda_i$ denotes any of the quartic couplings of the scalar potential. Because the singlet fermions ($N_i$) acquire their masses dynamically via $\langle\Phi\rangle$, i.e., $M_{N_i} = Y_{N_i} v_\phi/\sqrt{2}$, the Yukawa perturbativity limit establishes a lower bound on the vacuum expectation value:
\begin{equation}
  v_\phi \geq \frac{\sqrt{2}\, M_{N_3}}{\sqrt{4\pi}} \simeq 0.4\, M_{N_3}\,.
  \label{eq:vphibound}
\end{equation}
Additionally, scalar mixing induces the invisible decay $h \to JJ$, which is constrained by the experimental limit $\mathrm{BR}(h \to \mathrm{inv}) < 0.107$~\cite{ATLAS:2023tkt}. The inert doublet states are subjected to LEP searches, requiring $m_{\eta^\pm} > 100~\mathrm{GeV}$, $m_{\eta_R} + m_{\eta_I} > m_Z$, and $m_{\eta^\pm} + m_{\eta_{R,I}} > m_W$~\cite{ALEPH:2005ab,Pierce:2007ut,Lundstrom:2008ai}.

Subject to these bounds, we perform a comprehensive parameter scan over the ranges listed in Table~\ref{tab:scan}. In addition to these varying parameters, for simplicity, we fix $M_U=2~\mathrm{TeV}$ and $y_U=1$ throughout the scan. The neutrino Yukawa couplings $y_\nu$ are determined via the Casas-Ibarra parametrization~\cite{Casas:2001sr}, incorporating best-fit global neutrino oscillation data~\cite{deSalas:2020pgw} for normal ordering with a vanishing lightest neutrino mass ($m_{\nu_1} = 0$).
\begin{table}[ht]
  \centering
  \renewcommand{\arraystretch}{1.2}
  \begin{tabular}{ll}
    \toprule
  Parameter& \hspace{0.8cm}  Scan Range \\
    \midrule
 $M_{N_1}$                                      &\hspace{0.8cm} $200~\mathrm{GeV}$ - $5~\mathrm{TeV}$ \hspace{0.2cm}\\
    $M_{N_2}/M_{N_1}$                           &\hspace{0.8cm} $10^{0.1}$ - $10^{0.5}$ \\
    $M_{N_3}/M_{N_2}$                           &\hspace{0.8cm} $10^{0.05}$ - $10^{0.3}$ \\
    $v_\phi$                                    &\hspace{0.8cm} $\sqrt{2}\,M_{N_3}/\sqrt{4\pi}$ - $10^{4}~\mathrm{GeV}$ \\
    $m_J$                                       &\hspace{0.8cm} $0.22$ - $1.53~\mathrm{GeV}$ \\
    $\lambda_\eta$                              &\hspace{0.8cm} $5\times10^{-4}$ - $\sqrt{4\pi}$ \\
    $\lambda_\Phi$                              &\hspace{0.8cm} $5\times10^{-4}$ - $\sqrt{4\pi}$ \\
    $|\lambda_3|,\ |\lambda_4|$                 &\hspace{0.8cm} $10^{-3}$ - $\sqrt{4\pi}$ \\
    $\lambda_5$                                 &\hspace{0.8cm} $10^{-11}$ - $10^{-3}$ \\
    $|\lambda_{\eta\Phi}|$                      &\hspace{0.8cm} $10^{-3}$ - $1$ \\
    $|\lambda_{H\Phi}|$                         &\hspace{0.8cm} $10^{-5}$ - $0.1$ \\
    \bottomrule
  \end{tabular}
  \caption{Parameter ranges used in the numerical scan. All dimensionless couplings were sampled logarithmically. For simplicity, we fix $M_U=2~\mathrm{TeV}$ and $y_U=1$ throughout the scan.}
  \label{tab:scan}
\end{table}
The choice for the range of variation of the parameters listed in Table~\ref{tab:scan} is made keeping the above mentioned constraints in mind as well as to capture the range of interest for the LZ event.

\begin{figure}[ht]
  \centering
  \includegraphics[width=0.7\textwidth]{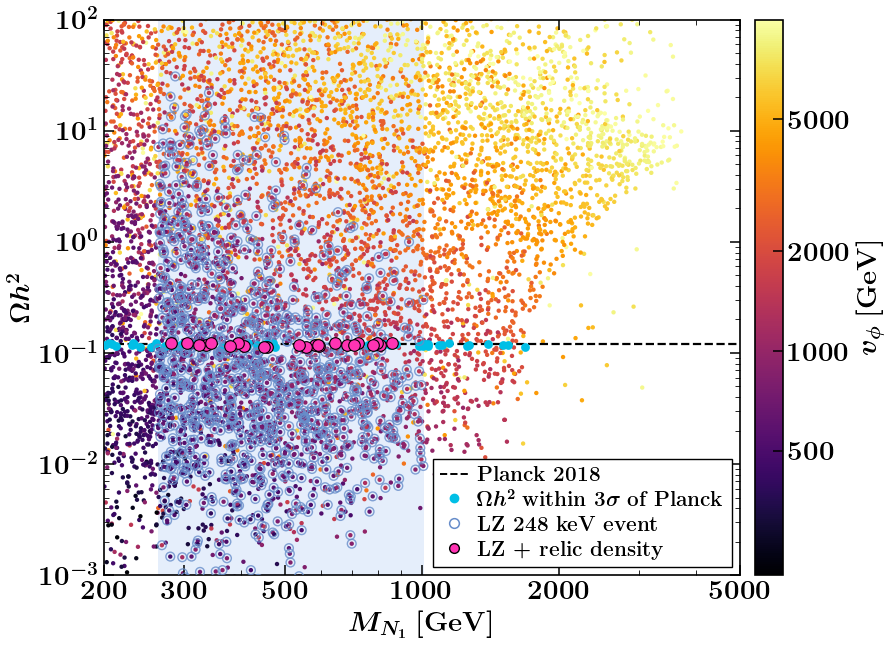}
  \caption{Relic abundance of dark matter candidate $N_1$ as a function of its mass $M_{N_1}$ for scan points with $v_\phi \leq 10~\mathrm{TeV}$, color coded by the symmetry breaking scale $v_\phi$. The dashed line shows the Planck 2018 central value $\Omega h^2 = 0.120$~\cite{Planck:2018vyg}. Cyan points reproduce the observed relic density within $3 \sigma$. Open blue circles denote points in the parameter region favoured by the LZ $248$~keV recoil event, whose mass range is indicated by the blue band, and pink points satisfy both conditions simultaneously.}
  \label{fig:relic}
\end{figure}

Fig.~\ref{fig:relic} displays the resulting dark matter $N_1$ relic abundance for scan points satisfying $v_\phi \leq 10~\mathrm{TeV}$. As discussed before, the collider bounds on the inert scalar masses typically restrict the parameter space only for masses below $M_{N_1} \sim 100~\mathrm{GeV}$. Because the region favored by the LZ recoil signal corresponds to heavier dark matter masses, these limits do not constrain the viable parameter space.

Additionally, charged lepton flavour violating (cLFV) processes, in particular $\mu \to e\gamma$, can place stringent constraints on parameter space of the model. However, as shown in older studies, they typically constraint only lighter dark matter mass well below our region of interest~\cite{Toma:2013zsa,Vicente:2014wga, CentellesChulia:2024iom}.

In contrast to the canonical scotogenic model, the annihilation here proceeds dominantly through the Majoron
channels, $N_1 N_1 \to JJ$ and $N_1 N_1 \to J\rho$, whose strength is governed by the Yukawa coupling
$Y_{N_1} = \sqrt{2}\, M_{N_1}/v_\phi$. Since $N_1$ acquires its mass from the vev of $\Phi$,
$Y_{N_1}$ grows as $v_\phi$ decreases for a given $M_{N_1}$.
A lower $v_\phi$ therefore enhances the Majoron mediated annihilation rate, leading to an
underabundant $N_1$, while a higher $v_\phi$ suppresses it and results in overabundance.
This is visible in Fig.~\ref{fig:relic} as a clear colour gradient. The purple points with small
$v_\phi$ lie at low $\Omega h^2$, while orange and yellow points with large $v_\phi$
populate the overabundant region. For each value of $M_{N_1}$, an appropriate $v_\phi$ reproduces
the observed relic density, so that a wide range of dark matter masses remains consistent
with the relic density constraint, bounded from above only by the perturbativity of the
Yukawa couplings $Y_{N_k}$.
We observe that the Majoron channels alone can
reproduce the observed relic density up to masses of a few TeV. The parameter region
favoured by the LZ $248$~keV recoil event lies at moderate values of $M_{N_1}$ and
$v_\phi$, and overlaps with the region where the observed relic abundance is obtained.
The pink points correspond to the dark matter masses satisfying all the relevant constraints while simultaneously  explaining the LZ event,
showing that the Majoron channels alone can account for both.

Thus, within the Majoron scotogenic model, Majoron mediated elastic DM-nucleus scattering provides a viable explanation for the high-recoil LZ event while remaining consistent with the observed cosmological dark matter relic density and other phenomenological constraints. Similar consistent realizations can also be expected in various variants of the Majoron scotogenic framework considered here, as well as in other suitable Majoron based neutrino mass models. Finally, we also expect the Dirac neutrino mass models with the pseudoscalar analogue of Majoron, often call Diracon, to also provide similar consistent realizations.

\section{Conclusions}

The connection between the LZ high-recoil event and radiative neutrino mass appears particularly compelling in scotogenic models, where an inelastic dark matter realization arises naturally. However, the mass splittings relevant to the event also allow efficient capture on heavy solar elements, placing these realizations under strong pressure from solar neutrino searches. We have explored whether this connection can be retained without relying on inelastic kinematics.

We instead pursue elastic scattering through a pseudoscalar mediator and embed it directly into the mechanism responsible for neutrino mass generation. The mediator is identified with the Majoron associated with spontaneous lepton number breaking, making the momentum dependent interaction required by the LZ spectrum a natural ingredient of the neutrino mass mechanism frameworks with spontaneously broken  lepton number $U(1)_L$ symmetry. Following the general model independent analysis at the effective operator level, we consider the scotogenic model as a specific UV complete framework. The lightest scotogenic singlet fermion provides Majorana dark matter, while a vector-like quark and a type-I Dirac seesaw transmit the Majoron interaction to the up quark.

This realization preserves the connection between the LZ event, dark matter and radiative neutrino mass while avoiding the solar capture problem faced by the inelastic scenario. The $\mathcal O_6$ interaction eliminates capture on the dominant zero nuclear spin solar isotopes and strongly suppresses capture on hydrogen through poor kinematic matching and small momentum transfer. At the same time, the model reproduces the high-recoil event without overpopulating the lower-energy bins and yields the observed thermal relic abundance. Pseudoscalar-mediated elastic scattering therefore reopens a neutrino mass motivated interpretation of the LZ event that would otherwise appear strongly constrained by the Sun. Similar consistent realizations can also be expected in various variants of the Majoron scotogenic framework and in other suitable Majoron-based neutrino mass models. More generally, Dirac neutrino mass models involving the pseudoscalar analogue of the Majoron, commonly referred to as the Diracon, are also expected to provide similar realizations.

\section*{Acknowledgements}
S.~C.~Chulia. acknowledges support from the Spanish grants PID2023-147306NB-I00, CNS2024-154524 and CEX2023-001292-S (MICIU/AEI/10.13039/501100011033). S.~C.~Chulia thanks IISER Bhopal for its hospitality during his stay, where this collaboration started.
We thank A. Majumdar and H. K. Prajapati for helpful discussions.\\

R.S. would like to dedicate this work to his toddler daughter, Akanksha, who bravely endured her father’s long absences while this work was in progress, without ever shedding a tear. 

\bibliographystyle{utphys2}
\bibliography{biby}

\end{document}